\documentclass[aps,prb,twocolumn,superscriptaddress,amsmath,amssymb]{revtex4-1}

\usepackage{graphicx}
\usepackage{dcolumn}
\usepackage{bm}

\begin{document}


\title{Insulator--to--Metal Transition in Selenium--Hyperdoped Silicon: Observation and Origin}


\author{Elif Ertekin}
\email{E. Ertekin (elif1@mit.edu) and M. T. Winkler (mwinkler@mit.edu) contributed equally to this work.}
\affiliation{Department of Materials Science and Engineering, and}

\author{Mark T. Winkler}
\email{E. Ertekin (elif1@mit.edu) and M. T. Winkler (mwinkler@mit.edu) contributed equally to this work.}
\affiliation{Department of Mechanical Engineering, Massachusetts Institute of Technology, Cambridge MA 02139}

\author{Daniel Recht}
\affiliation{Harvard School of Engineering and Applied Sciences, Cambridge MA 02139}

\author{Aurore J. Said}
\affiliation{Harvard School of Engineering and Applied Sciences, Cambridge MA 02139}

\author{Michael J. Aziz}
\affiliation{Harvard School of Engineering and Applied Sciences, Cambridge MA 02139}

\author{Tonio Buonassisi}
\affiliation{Department of Mechanical Engineering, Massachusetts Institute of Technology, Cambridge MA 02139}

\author{Jeffrey C. Grossman}
\email[e-mail: ]{jcg@mit.edu}
\affiliation{Department of Materials Science and Engineering, and}
\affiliation{Department of Mechanical Engineering, Massachusetts Institute of Technology, Cambridge MA 02139}


\date{\today}



\begin{abstract}
Hyperdoping has emerged as a promising method for designing semiconductors with unique optical and electronic properties, although such properties currently lack a clear microscopic explanation. Combining computational and experimental evidence, we probe the origin of sub--band gap optical absorption and metallicity in Se--hyperdoped Si. We show that sub--band gap absorption arises from direct defect--to--conduction band transitions rather than free carrier absorption. Density functional theory predicts the Se--induced insulator--to--metal transition arises from merging of defect and conduction bands, at a concentration in excellent agreement with experiment. Quantum Monte Carlo calculations confirm the critical concentration, demonstrate that correlation is important to describing the transition accurately, and suggest that it is a classic impurity--driven Mott transition. 
\end{abstract}
\maketitle  



%

Of all the experimentally measurable physical properties of materials, the electronic conductivity exhibits the largest variation, spanning a factor of 10$^{31}$ from the best metals to the strongest insulators~\cite{EdwardsRao}.  Over the last century, the puzzle of why some materials are conductors and others insulators, and the mechanisms underlying the transformation from one to the other, have been carefully scrutinized; yet even after such a vast body of research over such a long period, the subject remains the object of controversy.  In 1956, Mott introduced a model for the insulator-to-metal transition (IMT) in doped semiconductors, in which long--ranged electron correlations are the driving force~\cite{Mott56}.  Hyperdoping (doping beyond the solubility limit) creates a new materials playground to explore defect--mediated IMTs in semiconductors.  In this letter, we identify a defect--induced IMT in silicon hyperdoped with selenium concentrations exceeding $10^{20}$~cm$^{-3}$ (compared to the equilibrium solubility limit~\cite{Carlson59} of about $10^{16}$~cm$^{-3}$) and we explore the detailed nature of the transition with both experiment and computation.  We find that the IMT is largely driven by electron correlation and most resembles a classic impurity--driven Mott transition.  Additionally, we find that the high density of Se present at the IMT yields direct optical transitions and an absorption coefficient in excellent agreement with the measured sub-band gap optical properties \cite{Pan11}.

Hyperdoping is currently being used to engineer new materials with unique and exotic properties.  Silicon hyperdoped with chalcogens exhibits strong sub--band gap absorption down to photon energies as low as 0.5 eV~\cite{Younkin03,Crouch04,Sheehy05,Sheehy07,Kim06,Tabbal07,Wu01}, sparking substantial recent interest in applications such as infrared detection and intermediate band photovoltaics~\cite{Wu01,Younkin03,Crouch04,Sheehy05,Sheehy07,Kim06,Tabbal07}. The successful fabrication of rectifying junctions~\cite{Tabbal07} and photodiodes~\cite{Wu01,Carey05, Said11} using S and Se hyperdoped silicon appears to justify such interest.  While isolated S and Se dopants are well--established deep double donors in silicon~\cite{Carlson59,Grimmeiss81}, the enhanced optical properties of hyperdoped silicon (in which these chalcogenic impurities are present at much higher concentrations) are not yet well understood.  Further, unlike the prototypical system of phosphorus-doped silicon for which the IMT has been extensively studied and characterized~\citep{Rosenbaum83,Yamanouchi67}, there are very few studies of an IMT resulting from deep defects such as chalcogens~\citep{Winkler11}.

We prepared Se-doped silicon (Se:Si) samples using ion implantation followed by nanosecond pulsed--laser melting (PLM) and rapid resolidification. The PLM process enables chalcogen doping with concentrations exceeding 1\% atomic; such samples exhibit unexplained optical properties including broad, featureless absorption of photons with energy lower than the band gap of silicon~\cite{Kim06}. Silicon substrates (boron doped, $\rho \approx$ 25 $\Omega \cdot$cm) were ion implanted with Se to nominal doses of 3(10$^{15}$)  and 10$^{16}$ cm$^{л2}$ using an ion beam energy of 176 keV.  The implanted samples were exposed to four laser pulses (fluences of 1.7, 1.7, 1.7 and 1.8 J cm$^{-2}$).  This fluence regimen results in a slightly shallower dopant profile, and higher peak Se concentration, than reported previously\cite{Bob10}.  The Se--rich layer is crystalline, extends approximately 350 nm from the surface, and is electrically isolated from the p--type substrate by the rectifying junction formed between the two. The Se concentration-depth profile was measured via secondary ion mass spectrometry \cite{Bob10, Winkler11}. Sample preparation and measurement proceeded as described previously\cite{Winkler11}, with the Se:Si layer isolated using cloverleaf-mesa structures.   Conductivity was calculated from sheet conductivity using the effective implantation depth $d_{eff}$\cite{Winkler11}.

\begin{figure}
\includegraphics[width=8.8 cm]{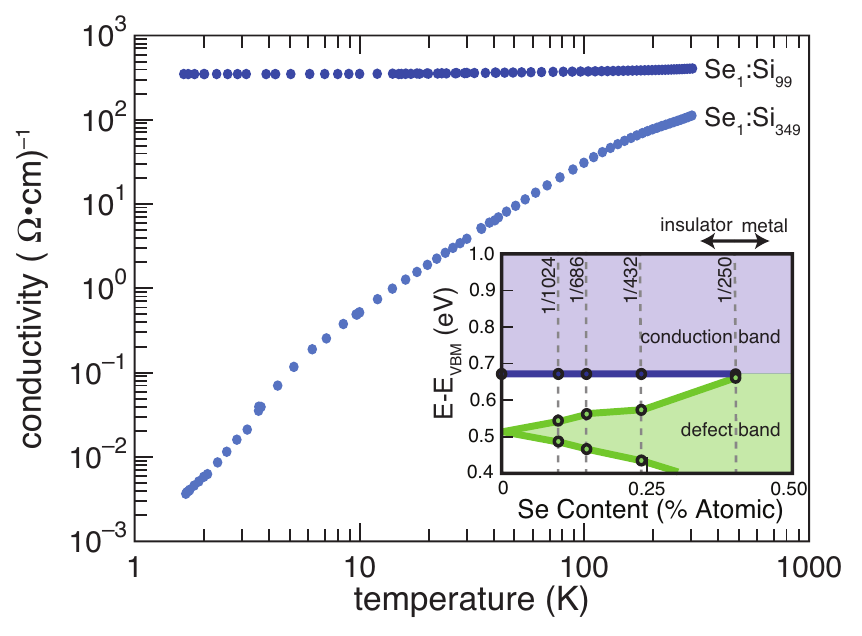}
\caption{\label{hall} {\bf Temperature--dependent conductivity of Se--hyperdoped silicon}.  At peak Se concentrations of Se$_1$:Si$_{349}$, hyperdoped Si exhibits strongly temperature--dependent conductivity, indicative of the insulating phase.  At higher concentrations (Se$_1$:Si$_{99}$), the conductivity is comparatively insensitive to temperature down to $T=1.8$~K, indicating a dopant--induced transition to the metallic state. {\bf Inset:} Analysis of density functional theory (DFT) calculations (shown in Fig. \ref{dos}) indicate that the conduction and defect bands cross as the Se--dopant concentration increases.  Conduction and defect band edges are demarked by DFT's Kohn--Sham eigenvalues for the insulating systems, with energies referenced to the valence band maximum $E_\mathrm{VBM}$.}
\end{figure}

\begin{figure*}
\includegraphics[width=17.0 cm]{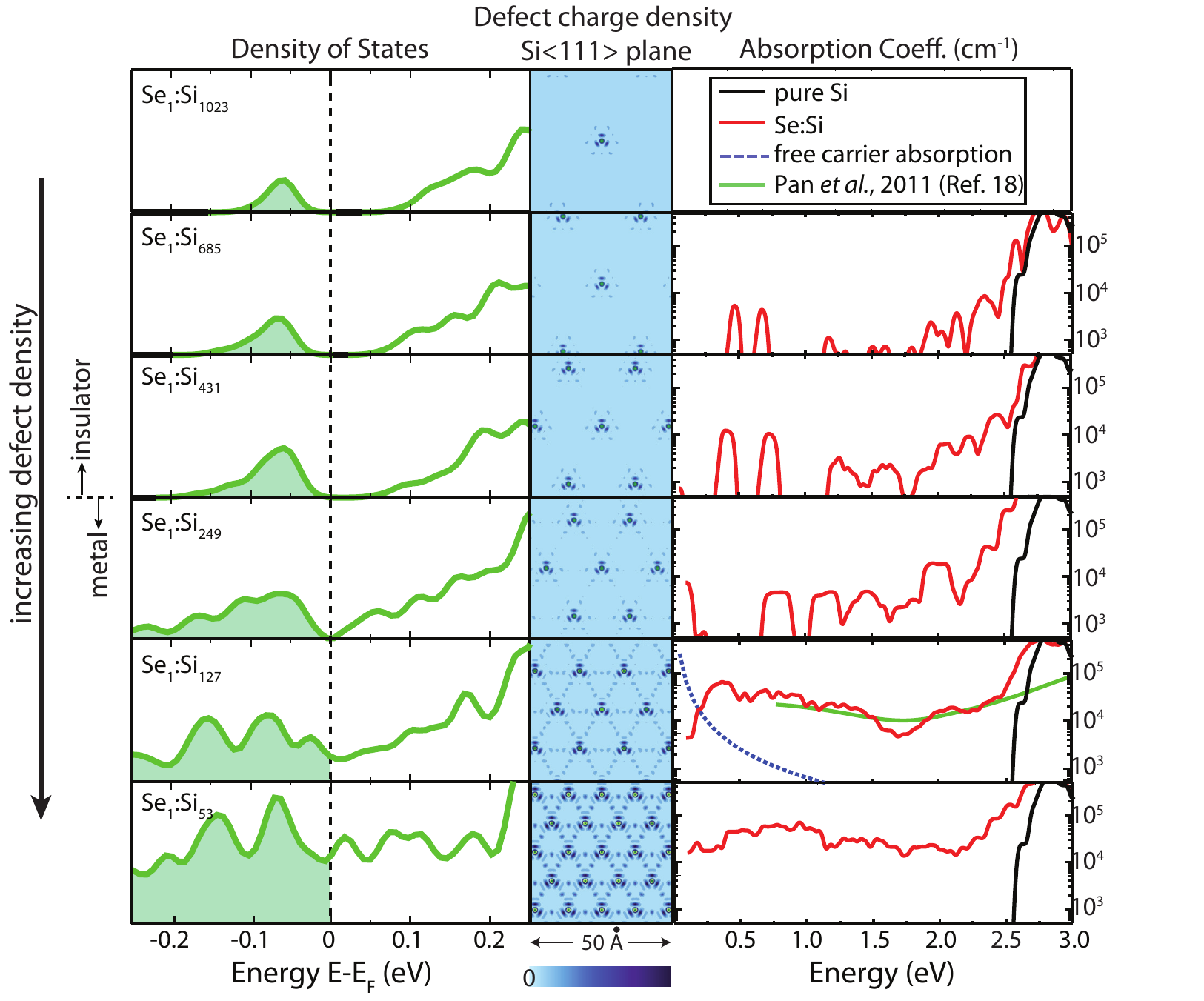}
\caption{\label{dos} {\bf Evolution of the insulator--to--metal transition in selenium--hyperdoped silicon.}  {\it Left column:} the electronic density of states (referenced to the Fermi energy $E_F$) for a range of Se concentrations, illustrating that the defect-- and conduction--bands merge at Se$_1$:Si$_{249}$.  The y--axis is uniform for all concentrations, and the DOS is normalized so that each plot contains the same number of atoms; shaded regions indicate occupied states at $T=0$~K.  {\it Middle column:}  charge density of the defect state at the $\Gamma$--point, plotted on the Si $\langle 111 \rangle$ plane.  Increasing dopant concentration increases interactions between neighboring defects, resulting in the eventual delocalization of the defect state.  {\it Right column:}  the calculated absorption coefficient arising from both (red) direct optical transitions (for pure and metallic Se:Si) and (blue) free carrier absorption (for metallic Se:Si), in comparison to pure silicon (black).  The green line for Se$_1$:Si$_{127}$ indicates to the experimental data of Ref.~[\onlinecite{Pan11}].}
\end{figure*}

Since charge carriers in the insulating phase require thermal activation, rigorous experimental proof of an IMT lies in the measurement of nonЛzero conductivity as the temperature $T$  approaches 0~K. In Fig.~\ref{hall}, we plot the temperature--dependent conductivity over a temperature range 1.8--300~K for two samples exhibiting peak Se concentrations of $1.4(10^{20})$ and $4.9(10^{20})$~cm$^{-3}$.  Despite the relatively small ($\sim$3.5x) difference in peak dopant concentration, the conductivity varies by almost 5 orders of magnitude at $T=1.8$~K. The dramatic difference in conductivity between two samples doped to similar levels, and the lack of significant temperature sensitivity as $T$ approaches 0~K in the more highly-doped sample, demonstrate a transition to metallic conduction at a concentration between that of the samples presented here.  For ease of subsequent comparison to computational results, we identify this concentration by the peak value of the of Se to Si ratio in these samples, thus between Se$_1$:Si$_{349}$ and Se$_1$:Si$_{99}$.

While the experimental evidence for an IMT is clear, it cannot indicate the origin of the transition.  The measured critical Se concentration is accessible, though, to both density functional theory (DFT) and the more accurate quantum Monte Carlo (QMC) methods. For shallow donors in silicon, in contrast, the transition occurs at roughly 1 dopant per 10000 atoms~\cite{Yamanouchi67}. The DFT~\cite{HohenbergKohn64,KohnSham65} results presented here employ the PBE approximation~\cite{PerdewBurkeErnzerhof96} to the exchange correlation functional as implemented within the SIESTA package~\cite{Artacho08}.  The inner core electrons are represented by Troullier--Martins pseudopotentials, and the Kohn--Sham orbitals are represented via a linear combination of numerical pseudo--atomic orbitals expanded in a triple--zeta with polarization gaussian basis set.  The DFT predicted lattice constants and band gaps are, respectively, for silicon 5.48 \AA (exp: 5.43 \AA) and 0.70 eV (exp: 1.1 eV), and for hexagonal selenium a=4.40 \AA, c/a=1.16 (exp: a=4.37 \AA; c/a=1.14) and 0.85 eV (exp: 1.8 eV). For each doping concentration, we substituted one silicon atom with a Se impurity and relaxed the atomic positions (so that atomic forces are $<$0.01 eV/\AA) and supercell lattice vectors (so that all stress tensor components are $<$2 kbar).  We chose the substitutional Se configuration, as our electronic structure calculations~\cite{NewmanTBS} as well as others~\cite{Mo04}  indicate the substitutional site to be the minimum energy defect configuration.  The QMC results presented here are computed via fixed node diffusion Monte Carlo conducted with the QWalk code~\cite{Wagner10}, with trial wave functions constructed with a Slater--Jastrow form using SIESTA's DFT orbitals, variance--minimized Jastrow coefficients, and a time step of 0.01 au.  The QMC energies are computed by averaging over twisted boundary conditions for all supercells.   Defect formation energies for both DFT and QMC were computed using $\Delta E_f = (E_{Se_1:Si_n}+\mu_{Si})-(E_{Si_{n+1}}+\mu_{Se})$,
wherein each atom's chemical potential $\mu$ is determined by the quasi--equilibrium of the doped silicon with SiSe$_2$ chains, which may be present as early stage precipitates~\cite{NewmanTBS}.

Using DFT, we first explore the electronic band structure, the electronic density of states (DOS), and the defect formation energy for supercells of size $n \times n \times n$, for $n=2,3,...,8$.  Using the 2--atom face--centered cubic unit cell, this sampling corresponds to systems of Se$_1$:Se$_{N-1}$, for $N=1024, 686, 432, 250, 128, 54,$ and $16$, and defect spacing that increases uniformly from one supercell to the next. The left--hand column of Fig.~\ref{dos} shows the computed DOS for all concentrations (shaded portions denote occupied states). At the lowest concentration (Se$_1$:Si$_{1023}$), an isolated defect peak of narrow bandwidth ($\sim$0.06~eV) is offset from the conduction band edge by $\sim$0.12~eV (estimated from the Kohn--Sham eigenvalues). This result is consistent with previous work~\cite{Sanchez10}. The defect band is completely occupied by the two extra electrons introduced by the Se impurity.  Because the filled defect band is offset from the conduction band, there are no nearby empty states and these extra electrons cannot contribute to metallic conduction. Thus, the system is in the insulating state.  The middle column of Fig.~\ref{dos} illustrates the charge density of the defect level, plotted at the $\Gamma$--point and on the Si $\langle 111 \rangle$ plane.  At the concentration of Se$_1$:Si$_{1023}$ the defect level charge density is highly localized around the impurity.

As the impurity concentration increases from Se$_1$:Si$_{1023}$ to Se$_1$:Si$_{249}$, the strongest effect evident in the DOS in Fig.~\ref{dos} is the increasing width of the defect peak, indicating increasing dispersion and interactions between neighboring defects (also apparent in the defect state charge densities in Fig.~\ref{dos}).  As defects become more closely spaced, stronger defect--defect interactions result in a decrease of the offset between the defect peak and the conduction band minimum in the DOS.  At the concentration of Se$_1$:Si$_{249}$ the defect peak just touches the conduction band edge, and at Se$_1$:Si$_{127}$ the defect and conduction bands have merged.  The disappearance of the offset between the defect peak and the conduction band edge at Se$_1$:Si$_{249}$ indicates that the dopant electrons are no longer bound to impurity sites.  Instead, many low--lying conduction--band like states are available for charge transport without thermal activation, signifying the onset of the IMT.  This band--crossing induced phase transition is also illustrated in Fig.~\ref{hall}. From Figs.~\ref{hall} (inset) and ~\ref{dos}, DFT predicts that the transition occurs at a concentration $\sim$Se$_1$:Si$_{249}$; this density is in excellent agreement with experiment.

For defect concentrations of Se$_1$:Si$_{127}$ and higher, the defect band further decomposes as it merges with the conduction band, as shown in Fig.~\ref{dos} and Fig.~\ref{hall}.  Large amounts of charge {\it between} impurities is observed in the charge densities of Fig.~\ref{dos}, indicating significant delocalization of the defect states.  The DFT analysis indicating hybridization of defect states with the conduction bands is consistent with experimental results, which indicate that the charge carriers exhibit conduction--band like character.  For example, we calculated the Hall mobility of the metallic sample in Fig.~\ref{hall} by measuring the Hall voltage as well as conductivity. The Hall mobility of the metallic sample is 21 $\pm$ 2~cm$^2$~V$^{-1}$~s$^{-1}$ at $T=4.2$ K, comparable in magnitude to that of metallic silicon doped with shallow donors (for which transport is known to occur in the conduction band)~\cite{Masetti,Yamanouchi67}. This high mobility value suggests that conduction in the metallic Se:Si sample arises from delocalization of Se electrons into the conduction band.  Thus both experiment and calculation support the idea that the IMT occurs due to a band--crossing of the Se defect states with the conduction band.  

\begin{figure}
\includegraphics[width=8.8 cm]{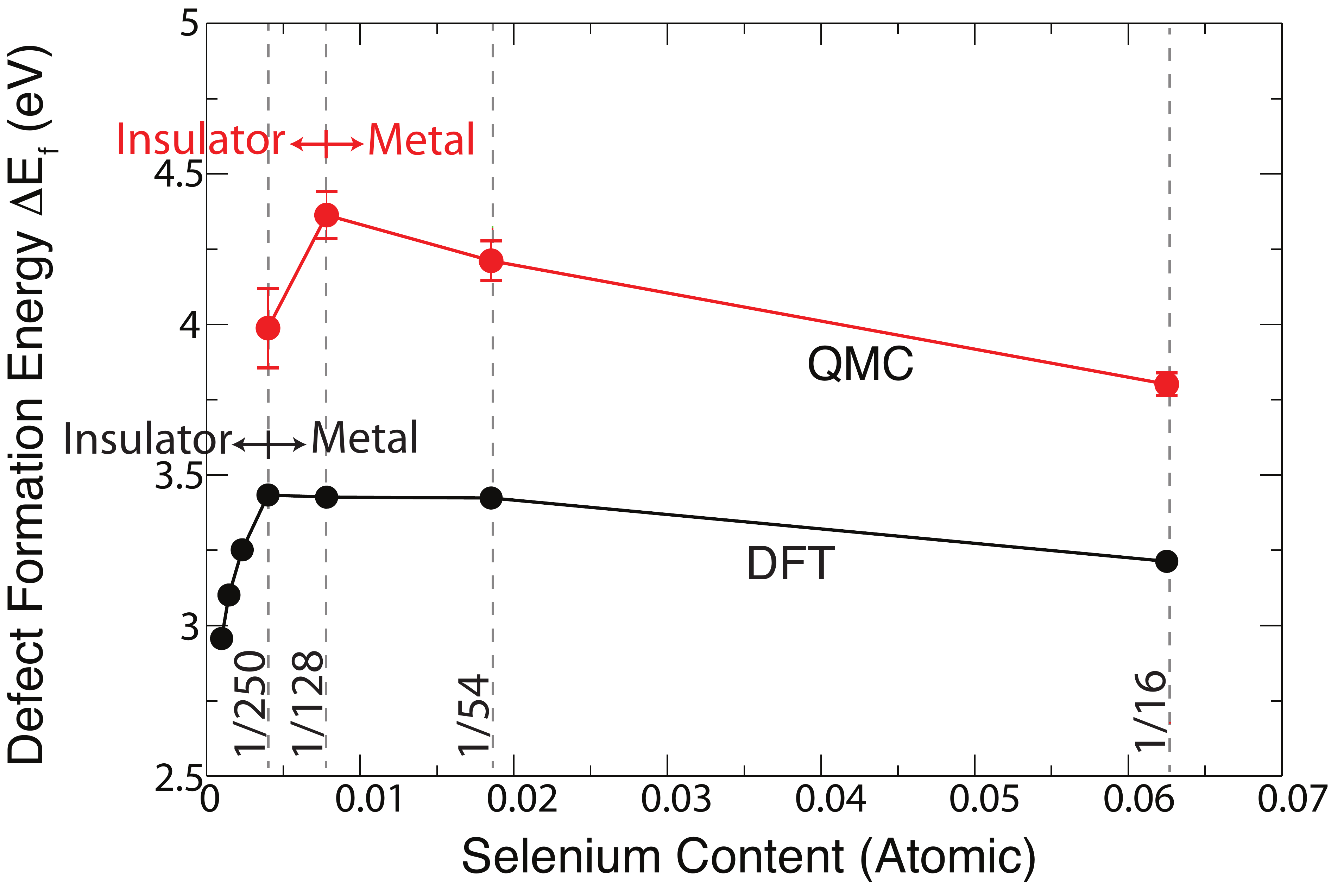}
\caption{\label{DefectEn} {\bf Defect formation energy vs. defect concentration}.  Formation energy of a substitutional Se dopant in Si,  computed via DFT and QMC.  According to DFT, a peak appears in the formation energy near the concentration Se$_1$:Si$_{249}$ corresponding to the IMT.   This peak is enhanced and shifted towards higher concentrations (Se$_1$:Si$_{127}$) in the QMC calculations, as is the phase transition.  Also, the QMC мcorrectionМ to DFT is largest near the transition, suggesting that electron correlation effects are prevalent at these concentrations.  }
\end{figure}

In the right--hand column of Fig.~\ref{dos}, we show contributions to the absorption coefficient arising from direct optical transitions calculated using our DFT results. In addition, contributions arising from free carrier absorption are shown, obtained using the Drude description and the experimentally measured free carrier concentration for the metallic Se:Si sample~\cite{Schroder78}.  Figure~\ref{dos} demonstrates that the defect levels predicted by our DFT calculations, and the resulting direct optical transitions and absorption coefficient, yield excellent agreement with both the magnitude and shape of the measured sub--band gap optical absorption in Se--hyperdoped silicon~\cite{Bob10}.  Free carrier absorption, alternatively, does not become comparably important except at photon energies lower than 200 meV.  

We now consider the underlying nature of the IMT itself.  In 1956, Mott introduced a model for an IMT in doped semiconductors driven by long--ranged electron correlations \cite{Mott56}, proposing that the delocalization of the electronic wave function occurs through long--ranged electron--electron interactions that screen the binding field of the impurity potential \cite{Mott56,Mott68}.  In the extensively studied semiconductor systems (such as P:Si), the measured critical concentration for metallic onset coincides well with concentration predicted by the Mott criterion\cite{Mott56}. Alternative IMT mechanisms have been identified\cite{EdwardsRao}, including transitions arising from static disorder (Anderson mechanism), local electron--electron correlations (Mott--Hubbard mechanism), and strong electron--phonon coupling; in real systems, more than one mechanism may contribute simultaneously. Definitive proof that the observed IMT is driven by long--ranged electron correlation is difficult, but QMC can establish the relative importance of electron correlations on the defect formation energies in both insulating and metallic systems.  DFT treats electron correlation in an approximate manner (correlation refers here to contributions to the total energy beyond the independent electron approximation, including the exchange contribution), and describes total energies better for metallic systems than insulating ones.  QMC, alternatively, provides an accurate description of electron correlation\cite{Grossman01,Wagner10}, and can accurately describe both states.  We look for clues to the origin of the IMT in the DFT and QMC defect formation energies.  

In Fig.~\ref{DefectEn}, the formation energy $\Delta E_f$ of a Se defect is shown as a function of defect concentration. An indication of a phase transition will appear as a kink (discontinuity in the first or higher derivative) in the curve showing the total energy per atom vs. dopant concentration; or equivalently as a kink in the curve showing defect formation energy vs. dopant concentration (since one is an affine transformation of the other).  All experimentally known transitions in semiconductors currently appear to be continuous (notably a first--order Mott transition in LiCoO$_2$ is believed to exist~\cite{Marianetti04}).  In DFT, the defect formation energy peaks in the vicinity of the IMT, although the discrete sampling renders it difficult to determine the order of any potential discontinuity.  On the insulating side, the defect formation energy increases with defect concentration, likely because correlation energy (a stabilizing contribution) tends to decrease with increasing electron density.  Thus the penalty for assigning additional Se atoms becomes more costly and ultimately renders the insulating state unstable.  At higher Se concentrations than the IMT, the defect formation energy is relatively stable, slightly decreasing with increasing defect concentration.  Using QMC (which, due to its computational cost was only performed for the Se$_1$:Si$_{15}$, Se$_1$:Si$_{53}$, Se$_1$:Si$_{127}$, and Se$_1$:Si$_{249}$, samples), the peak in $\Delta E_f$ is amplified and shifted towards Se$_1$:Si$_{127}$. This result remains in agreement with experiment, although QMC calculations should more accurately predict the transition point than DFT, which tends to overly delocalize electronic states.  Also, the QMC мcorrectionМ to the defect formation energy is larger in the vicinity of the phase transition than for the metallic systems, indicating that as expected DFT more accurately simulates the metal than the insulator.  This result suggests that many body effects are particularly important near the transition point and --- although not a rigorous proof that the phase transition is driven by correlation --- indicates that the IMT in Se:Si exhibits a strong Mott-like component.  The differences between DFT and QMC descriptions of the Se:Si system clearly illustrate the importance of accurately treating electron correlation in the fundamental study of phase transitions.

Finally, we comment on the implications of our analysis on the technological applications of silicon hyperdoped with chalcogens, which has been considered as a candidate material for infrared absorbers and intermediate band photovoltaics (IBPVs).  The anomalous sub--band gap absorption observed in hyperdoped samples can be understood in terms of the electronic band structure diagrams (showing a highly dispersive defect band that merges with the conduction bands) and the calculated absorption spectra.  We believe that Se:Si is an intriguing candidate for an infrared detector, exhibiting strong absorption down to low photon energies.  Although the experimentally realized system may differ in important ways, the system we model computationally would perform poorly as an IBPV due to the lack of unoccupied states in the intermediate band, in agreement with Ref.~\onlinecite{Sanchez10}.  
We note, however, that we have considered an uncompensated system in the above discussion, and that compensation may permit additional degrees of control over the properties of hyperdoped materials.  We are optimistic, though, that we can use the framework described in this Letter to predict and screen new candidates for intermediate band properties. \\

\noindent {\bf Acknowledgements}
{\small
EE and JCG are supported by DOE grant DE--SC0002623.  MTW and TB's work was supported by the U.S. Army Research Laboratory and the U.S. Army Research Office under grant number W911NF-10-1-0442.  DR, AJS, and MJA are supported by grant U.S. Army--ARDEC under contract W15QKN-07-P-0092.  Calculations were performed in part at the National Energy Research Scientific Computing Center of the Lawrence Berkeley National Laboratory and in part by the National Science Foundation through TeraGrid resources provided by NCSA under grant TG-DMR090027.   The authors acknowledge Jacob Krich, Christie Simmons, Bonna Newman, Meng--Ju Sher, Joseph Sullivan, and Lucas Wagner for insightful discussions and Tom Mates (under NSF contract DMR 04-20415) for the SIMS measurements.  
} \\

\bibliography{apstemplate}

\end{document}